\documentclass[reprint,amsmath,amssymb,aps,nofootinbib,superscriptaddress]{revtex4-2}

\usepackage{graphicx}
\usepackage{bm}
\usepackage{color}

\begin{document}

\preprint{APS/123-QED}

\title{Unveiling orbital landscapes in strongly correlated bulk nickelates with $s$-NIXS}

\author{Edgar Abarca Morales}
\email{Edgar.Morales@cpfs.mpg.de}
\affiliation{Max Planck Institute for Chemical Physics of Solids, N\"othnitzer Strasse 40, 01187 Dresden, Germany}

\author{Martin Sundermann}
\affiliation{Max Planck Institute for Chemical Physics of Solids, N\"othnitzer Strasse 40, 01187 Dresden, Germany}
\affiliation{Deutsches Elektronen-Synchrotron (DESY), Notkestraße 85, 22607 Hamburg, Germany}

\author{Brett Leedahl}
\affiliation{Max Planck Institute for Chemical Physics of Solids, N\"othnitzer Strasse 40, 01187 Dresden, Germany}

\author{Vignesh Sundaramurthy}
\affiliation{Max Planck Institute for Solid State Research, Heisenbergstraße 1, 70569 Stuttgart, Germany}

\author{Georg Poelchen}
\affiliation{Max Planck Institute for Chemical Physics of Solids, N\"othnitzer Strasse 40, 01187 Dresden, Germany}

\author{Ulrich Burkhardt}
\affiliation{Max Planck Institute for Chemical Physics of Solids, N\"othnitzer Strasse 40, 01187 Dresden, Germany}

\author{Raul Cardoso}
\affiliation{Max Planck Institute for Chemical Physics of Solids, N\"othnitzer Strasse 40, 01187 Dresden, Germany}

\author{Pascal Puphal}
\affiliation{Max Planck Institute for Solid State Research, Heisenbergstraße 1, 70569 Stuttgart, Germany}

\author{Alexander Komarek}
\affiliation{Max Planck Institute for Chemical Physics of Solids, N\"othnitzer Strasse 40, 01187 Dresden, Germany}

\author{Bernhard Keimer}
\affiliation{Max Planck Institute for Solid State Research, Heisenbergstraße 1, 70569 Stuttgart, Germany}

\author{Matthias Hepting}
\affiliation{Max Planck Institute for Solid State Research, Heisenbergstraße 1, 70569 Stuttgart, Germany}

\author{Liu Hao Tjeng}
\affiliation{Max Planck Institute for Chemical Physics of Solids, N\"othnitzer Strasse 40, 01187 Dresden, Germany}

\author{Berit H. Goodge}
\email{Berit.Goodge@cpfs.mpg.de}
\affiliation{Max Planck Institute for Chemical Physics of Solids, N\"othnitzer Strasse 40, 01187 Dresden, Germany}

\date{\today}

\begin{abstract}
We leverage $s$-orbital non-resonant inelastic X-ray scattering ($s$-NIXS) to perform orbital imaging on three bulk rare-earth nickelates spanning a range of formal nickel valence (3$d$ electron filling) from Ni$^{3+}$ (3$d^7$) to Ni$^{1+}$ (3$d^9$). Our results directly reveal the ground states of these compounds all with minimal theoretical input. In particular, we demonstrate the low-spin orbital configuration of trivalent LaNiO$_3$, the $d_{x^2-y^2}$ configuration of monovalent LaNiO$_2$, and resolve the effective $e_g$ crystal field splitting in the distorted octahedral environment of divalent La$_2$NiO$_4$. This work illustrates the potential of $s$-NIXS to study the ground state and excited states of strongly correlated materials without needing complex theoretical analysis of spectroscopic data.
\end{abstract}

\maketitle

A rich interplay between spin, charge, and orbital degrees of freedom gives rise to the exotic and functional properties in quantum and strongly correlated materials \cite{khomskii2014transition,tokura2017emergent}. Atomic-like $d$ orbital states in particular are often considered the drivers of such phenomena as spin- and charge-ordering, unconventional superconductivity, magnetism, and multiferroicity. Understanding and controlling these states and their interactions is therefore a primary goal of many thrusts in condensed matter physics on the road to bespoke functional materials by design. This calls for the development and expansion of advanced probes, especially for the orbital degrees of freedom in electronic structure.

\begin{figure*}
\includegraphics{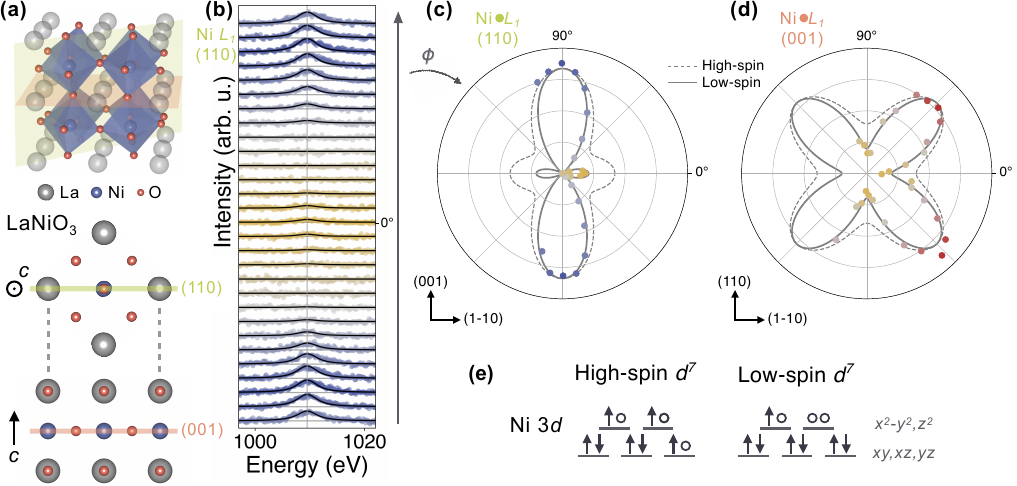}
\caption{\label{fig:fig1} Orbital imaging in LaNiO$_3$. (a) Pseudo-cubic unit cell of LaNiO$_3$ and two high-symmetry projections in which octahedral distortions are omitted for clarity. The $(110)$ and $(001)$ scattering planes are indicated in green and salmon respectively. (b) Ni $L_1$ edge spectra measured within the $(110)$ plane and fitted using Voigt distributions (black lines). Spectra are vertically offset by $\phi$ for clarity. (c) Polar plot of the integrated intensities from the Voigt fits in (b). The expected orbital hole shapes for the high- and low-spin configurations are indicated by dashed and solid gray lines, respectively. (d) Polar plot obtained from the analysis of the spectra within the $(001)$ plane (Supplemental Fig.~S1). (e) Schematic one-electron energy level diagram for the different Ni $3d$ ground states.}
\end{figure*}

Techniques such as photoelectron spectroscopy, scanning tunneling microscopy, resonant inelastic X-ray scattering (RIXS), X-ray absorption spectroscopy (XAS), and electron energy loss spectroscopy (EELS) have proven invaluable for disentangling the electronic behavior of complex systems. Probing the $d$ orbitals which are involved in the formation of the ground state and its low-energy excitations, for example, has so far been primarily achieved through comparison between spectroscopic measurements with modelling. In XAS, for instance, polarization-dependent absorption based on linear dichroism has been employed to visualize the orbital polarization of materials \cite{chen1992out,park2000spin,haverkort2005orbital,csiszar2005controlling}. This approach then relies on reproducing the experimental spectra with atomic multiplet and cluster configuration interaction calculations, which can be rather elaborate and require careful tuning of several semi-empirical parameters \cite{tanaka1994resonant,degroot1994xray,haverkort2012multiplet}. Such calculations further become quickly very complex if more than one correlated site must be included to describe the system, such as in materials where the strongly correlated $d$ electrons form bands that are part of the Fermi surface.

Consequently, expanding the experimental toolkit with complementary techniques is essential to achieve a better understanding of the electronic and magnetic properties of complex correlated materials. Recently, $s$-orbital non-resonant inelastic X-ray scattering ($s$-NIXS) was developed \cite{yavacs2019direct, leedahl2019origin, amorese2021selective} to obtain information on the orbital occupation and their excitations by making direct images of the orbitals involved, without the need to do many-body calculations. This element-specific, photon-in photon-out spectroscopy operates at high momentum transfer, where a hard X-ray beam of approximately 10 keV is scattered from the sample in a fixed $2\theta$ geometry, and the energy loss of the outgoing photons is analyzed. The use of high momentum transfer, achieved here by setting $2\theta = 155^\circ$, is required to access quadrupolar $s\rightarrow{d}$ transitions which are dipole-forbidden in many standard spectroscopic techniques such as XAS, RIXS, or forward-scattered EELS. Keeping the scattering geometry fixed, the experiment measures the spectra of single crystals as a function of the sample orientation. Because the initial $s$ core level state is spherically symmetric, the orientational dependence of integrated spectral intensity provides direct information on the symmetry of the ground state or initial state, here the $d$-hole charge density. In addition, the study of the excited states is facilitated by the orbital images that can be observed in the orientational dependence of the various spectral features. 

Here, we utilize the power of $s$-NIXS to probe electronic configurations across the family of strongly correlated rare-earth nickelates, which host a rich variety of quantum phases and electronic structures in their perovskite, reduced, and layered forms \cite{torrance1992systematic,catalan2008progress,zhang2017large,li2019superconductivity,pan2022superconductivity,sun2023signatures,li2024signature, ko2025signatures}. Specifically, we investigate the many-body ground states of three bulk compounds, namely LaNiO$_3$ \cite{sreedhar1992electronic}, La$_2$NiO$_4$\cite{rodriguez1991neutron}, and LaNiO$_2$ \cite{hepting2020electronic}, which together span formal Ni valence from 3+ to 1+. Details of sample synthesis are provided in Methods. The $s$-NIXS measurements in this work were performed at the High-Resolution Dynamics Beamline P01 of PETRA-III synchrotron in Hamburg, Germany.

\section{Perovskite LaNiO$_3$}
 
The rare-earth perovskite nickelates RNiO$_3$ (R = rare earth) have attracted significant and sustained research attention over several decades for both fundamental and applied interests \cite{medarde1997structural, catalan2008progress, middey2016physics}. The most notable functional property of perovskite nickelates is a sharp metal-insulator transition (MIT) which can be tuned over an impressive temperature range through choice of rare-earth ion or their mixing, epitaxial strain, or externally applied forces \cite{torrance1992systematic,catalan2008progress, middey2016physics, catalano2018rare,obradors1993pressure, canfield1993extraordinary}, which has been proposed as a platform for potential applications ranging from neuromorphic computing to optical camouflage to smart windows \cite{zhang2022reconfigurable, li2016correlated, sun2021electrochromic}. The MIT in many RNiO$_3$ compounds is accompanied by charge and magnetic ordering. From a more fundamental perspective, perovskite nickelates thus offer a rich playground for investigating the close link between lattice structure, electron correlations, and material properties. Despite the relative simplicity of their crystal structure, the ground-state electronic configuration of perovskite nickelates was discussed for many years to establish the appropriate orbital configuration for the unusually high formal valence of Ni$^{3+}$. Here, we investigate the first member of the perovskite nickelate family, LaNiO$_3$. It is the only member which does not exhibit an MIT, and serves as a prototypical example of the metallic state. 

The nominal Ni$^{3+}$ ion resides in a regular ($O_h$) octahedral coordination, which exhibits tilts and rotations within a pseudo-cubic unit cell (see Fig.~\ref{fig:fig1}a). For clarity, the subtle octahedral distortions of LaNiO$_3$ have been omitted in the projections at the bottom of Fig.~\ref{fig:fig1}a, as $s$-NIXS averages over many unit cells, effectively canceling such distortions which occur symmetrically around the Ni site. Two scattering planes, $(001)$ and $(110)$, were chosen to probe the orbital symmetry at the Ni site. Fig.~\ref{fig:fig1}b shows the $s$-NIXS spectra corresponding to the $2s\rightarrow{3d}$ transition measured at the Ni $L_1$ edge, obtained while rotating the sample by an angle $\phi$ within the $(110)$ scattering plane. For each acquisition, the energy scale was calibrated using the elastic peak and spectra were normalized to the Compton profile following the procedure described in \cite{yavacs2019direct}. 

A linear function fit to the pre- and post-edge region was used to subtract the background. A clear angular dependence of the spectral intensity is observed for the single peak centered around 1009.5 eV, here fitted with Voigt distributions that account for the experimental resolution of 1.5 eV (Gaussian) and a core-hole lifetime broadening of approximately 4 eV (Lorentzian). Integrated intensities of the fitted Voigts in Fig.~\ref{fig:fig1}b are plotted as a function of $\phi$ in Fig.~\ref{fig:fig1}c. Similar analysis was performed for the $(001)$ scattering plane, with the corresponding results shown in Fig.~\ref{fig:fig1}d. The complete set of $s$-NIXS spectra in this work is provided in Supplemental Note A (Figs.~S1-S3), details on the fitting procedures are given in Supplemental Note B (Figs.~S4-S7), and the Ni peak positions are summarized in the table in Supplemental Note C. 

\begin{figure*}[t!]
\includegraphics{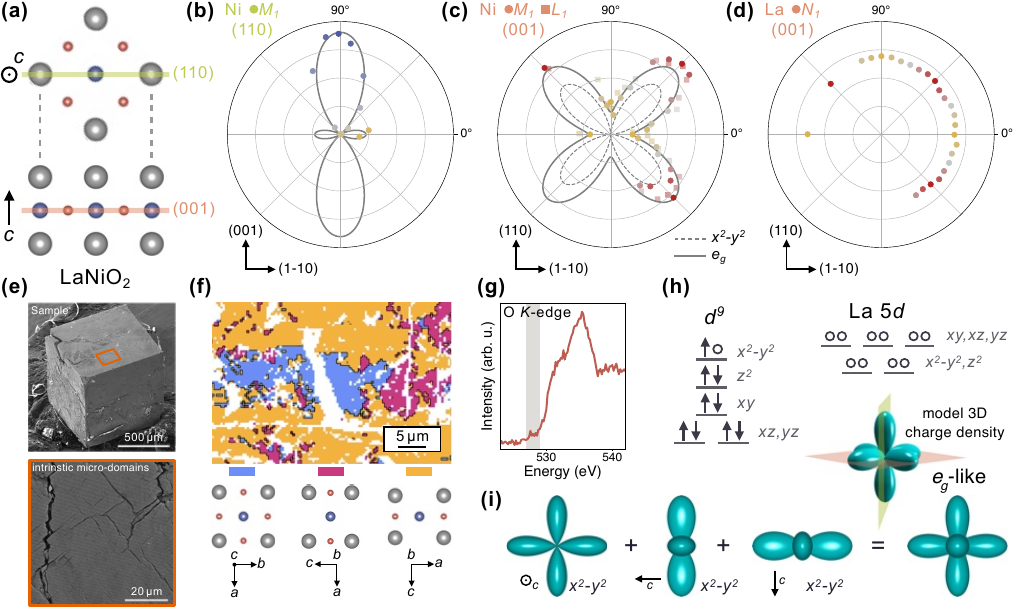}
\caption{\label{fig:fig2} Orbital imaging in LaNiO$_2$. (a) Crystal structure of LaNiO$_2$ with the $(001)$ and $(110)$ scattering planes indicated. (b) Polar plot obtained from the Ni $M_1$ edge spectra within the $(110)$ plane. (c) Polar plots from the Ni $M_1$ edge and $L_1$ edge spectra within the $(001)$ plane. The $e_g$ and $d_{x^2-y^2}$ orbital shapes are indicated by gray lines. (d) Polar plot from the La $N_1$ edge spectra within the $(001)$ plane. (e) Bulk LaNiO$_2$ sample with inset showing intrinsic micro-domain formation. (f) EBSD map of LaNiO$_2$ showing three orthogonal domains and their schematic microscopic configuration (adapted from \cite{hayashida2024investigation}). (g) O $K$-edge XAS of LaNiO$_2$ a lack of the O pre-peak (region highlighted in gray). (h) Schematic electronic configurations for Ni$^{1+}$ $3d^9$ and La$^{3+}$ $5d^0$ states. Note that in the canonical square-planar ligand field, the $xy$ states are at higher energy than the $z^2$; here we present the configuration adjusted for the calculated band alignment in infinite-layer nickelates. (i) Sum of three orthogonally-oriented $d_{x^2-y^2}$ charge densities yielding an $e_g$-like orbital shape. The $(001)$ and $(110)$ scattering planes are indicated in the 3D visualization.}
\end{figure*}

The results in Figs.~\ref{fig:fig1}c~and d clearly indicate an $e_g$-like ($d_{x^2-y^2}+d_{z^2}$) hole density at the Ni site, similar to that observed in NiO with a canonical Ni $d^8$ configuration \cite{yavacs2019direct}. These measurements definitively exclude the possibility of the presence of a hole in the $t_{2g}$ subshell since this would result in broadened polar features as shown by the dashed line in Figs.~\ref{fig:fig1}c~and d. We thus can rule out the high-spin 3$d^7$ $t_{2g}^5$$e_g^2$ configuration (Fig.~\ref{fig:fig1}e), leaving us with the low-spin 3$d^7$ $t_{2g}^6$$e_g^1$ or $3d^8\underline{L}$ $t_{2g}^6$$e_g^2$$\underline{L}_{e_g}$ configurations, where $\underline{L}$ indicates a hole in the O $2p$ ligand; intersite pair combinations of $3d^8\underline{L}^2$ and $3d^6$ with the $t_{2g}$ subshell filled are also possible \cite{green2016bond, bisogni2016ground}. 

Previous efforts to identify the ground state configuration in RNiO$_3$ compounds relied on comparison between experimental measurements with detailed charge transfer multiplet or cluster calculations, for example to estimate fractional contributions from low-spin and high-spin configurations \cite{piamonteze2005spin}, or to infer charge transfer and hybridization with continuum or band states \cite{freeland2016evolution}. Even more complex calculations have been performed on two-site cluster calculations to include inter-site charge transfer and hybridization effects \cite{green2016bond, bisogni2016ground}. Instead, $s$-NIXS provides a direct experimental probe of the electronic ground state charge density \emph{without} any theoretical input or calculations, highlighting the potential of extending this technique to other strongly correlated materials where the $d$ electrons form bands that cross the Fermi level.
The remaining challenge is to explain the paramagnetic behavior of LaNiO$_3$ over a wide range of temperatures together with the mass enhancement \cite{zhou2014mass} relative to the density functional results. The development of such a many-body model requires a mechanism that screens out local spin moments. 

\section{Infinite-layer LaNiO$_2$}

Scientific interest in rare-earth nickelates further expanded in 2019 with the discovery of superconductivity in hole-doped infinite-layer nickelates RNiO$_2$ \cite{li2019superconductivity}. Originally proposed as possible cuprate analogues \cite{anisimov1999electronic}, experimental and theoretical efforts since their synthetic stabilization have largely focussed on understanding the detailed electronic and magnetic properties of these compounds in connection to cuprates and other superconducting families \cite{nomura2022superconductivity, wang2024experimental, puphal2026superconductivity}. The infinite-layer crystal structure can be considered a derivative of the perovskite structure with the apical oxygens in the R plane removed, as illustrated in the projections of Fig.~\ref{fig:fig2}a. The absence of these oxygen atoms induces a square-planar coordination around the Ni site, resulting in a nominal Ni$^{1+}$ metastable configuration within the tetragonal unit cell. 

Orbital imaging of the Ni site within the $(001)$ and $(110)$ scattering planes is shown in Figs.~\ref{fig:fig2}b and~\ref{fig:fig2}c, respectively. Here, we focus on spectra acquired from the $3s \rightarrow 3d$ transition at the Ni $M_1$ edge, which offers the advantage of being more intense and less broadened than the $L_1$ edge, though at the cost of being relatively close in energy to the La $N_{4,5}$ multiplet and its EXAFS tail (Supplemental Fig. S2). Importantly, both edges provide equivalent physical information, as they originate from $s$-orbital initial states and correspond to $3d$ final states. This is illustrated in Fig.~\ref{fig:fig2}c which shows the combined normalized polar intensities for both Ni $M_1$ and Ni $L_1$ spectra.

The orbital symmetry at the La site within the $(001)$ plane is also shown in Fig.~\ref{fig:fig2}d for the $4s \rightarrow 5d$ transition at the La $N_1$ edge. The role of electron-like La 5$d$ states at or near the Fermi level which may contribute to self-doping effects in chemically undoped nickelates remains under discussion \cite{ding2024cuprate,sun2025electronic}.
Here, we find a rather featureless hole density which can be attributed to a fully unoccupied $5d$ manifold (Fig.~\ref{fig:fig2}h) within the sensitivity of our experiments. 

Similar to LaNiO$_3$, Figs.~\ref{fig:fig2}b and~\ref{fig:fig2}c reveal an $e_g$-like orbital symmetry at the Ni site. 
As illustrated in Fig.~\ref{fig:fig2}h, however, a $d_{x^2-y^2}$ hole density is expected for Ni$^{1+}$ (3$d^9$) in a canonical square-planar configuration. The most obvious distinction is made by the presence or lack of hole density in the (110) cut, which should remain nearly empty for a pure $d_{x^2-y^2}$ configuration, counter to the clear structure seen in Fig.~\ref{fig:fig2}b.
A more subtle but equally convincing distinction is also observed in Fig.~\ref{fig:fig2}c, where the $e_g$ (solid grey line) and $d_{x^2-y^2}$ (dashed line) hole densities can be clearly distinguished by the lack or presence respectively of zero-intensity nodes between their lobes. 

This apparent discrepancy between the expected $d_{x^2-y^2}$ and the observed $e_g$-like orbital shape can be naturally explained by the presence of orthogonal domains in the sample. In practice, samples are grown first as perovskite LaNiO$_3$ before the oxygen removal is achieved chemically to reach LaNiO$_2$ \cite{puphal2023synthesis}. In bulk crystals, this reduction process does not favor a specific direction of apical oxygen removal, so intrinsic microdomains form in the sample (see Fig.~\ref{fig:fig2}e) corresponding to the three orthogonal directions of oxygen depletion \cite{hayashida2024investigation,wu2024unraveling}, or equivalently to three orthogonal orientations of LaNiO$_2$. Electron backscatter diffraction (EBSD) measurements shown in Fig.~\ref{fig:fig2}f confirm the presence of these orthogonal domains, which are distributed roughly homogeneously throughout the sample \cite{hayashida2024investigation}. Importantly, oxygen $K$ edge XAS measurements demonstrate the complete reduction of the sample from Ni$^{3+}$ to Ni$^{1+}$, as evidenced by the absence of a prominent O pre-peak near 528 eV in Fig.~\ref{fig:fig2}g \cite{hepting2020electronic,goodge2021doping}. The $e_g$-like orbital density is therefore simply explained as the sum of $d_{x^2-y^2}$ charge densities along orthogonal directions, which is mathematically equivalent to an $e_g$-like orbital symmetry as depicted in Fig.~\ref{fig:fig2}i.

\begin{figure*}[t!]
\includegraphics{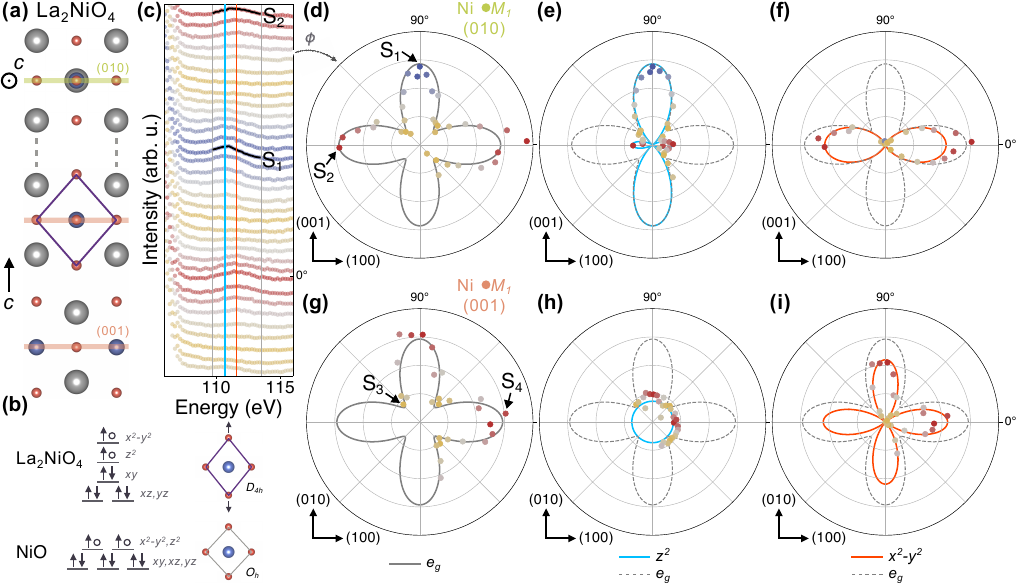}
\caption{\label{fig:fig3} Orbital imaging in La$_2$NiO$_4$. (a) Crystal structure of La$_2$NiO$_4$ indicating the $(010)$ and $(001)$ scattering planes as well as the octahedral coordination around the Ni site. (b) Schematic electronic configurations for the Ni ground state in La$_2$NiO$_4$ and NiO in a single-electron picture, illustrating the $e_g$ splitting due to the apical octahedral distortion in La$_2$NiO$_4$.(c) Ni $M_1$ edge spectra measured within the $(010)$ plane, vertically offset by $\phi$. The spectra selected to form the basis for orbital decomposition are marked in black, labeled $S_1$ and $S_2$. From this analysis, a $d_{z^2}$ component is identified at the blue line, while a $d_{x^2-y^2}$ component is found at the red line. (d) Polar plot of the integrated intensities of the spectra enclosed by the gray boundaries in (c). The black arrows indicate the data points for the labeled spectra in (c). The $e_g$ orbital symmetry is indicated by the gray line. (e) Polar plot of the $d_{z^2}$ orbital content along the $(010)$ plane. The $d_{z^2}$ orbital symmetry is indicated by the blue line. (f) Polar plot of the $d_{x^2-y^2}$ orbital content along the $(010)$ plane. The $d_{x^2-y^2}$ orbital symmetry is indicated by the red line. (g) Polar plot obtained from the Ni $M_1$ edge spectra within the (001) plane. The selected basis states $S_3$ and $S_4$ for the orbital decomposition are indicated (see Supplemental Note B). (h) Polar plot of the $d_{z^2}$ orbital content along the $(001)$ plane. (i) Polar plot of the $d_{x^2-y^2}$ orbital content along the $(001)$ plane. }
\end{figure*}

\section{Ruddlesden-Popper La$_2$NiO$_4$}

Instead of removing oxygen, the perovskite motif can also be modified through stacking confined layers of RNiO$_3$ separated by RO rock salt spacer layers to form the Ruddlesden-Popper series R$_{n+1}$Ni$_{n}$O$_{3n+1}$ \cite{rodriguez1991neutron,zhang2020high, pan2022synthesis}, where RNiO$_3$ is the $n = \infty$ endmember.
In nickelates, this layering provides a tuning knob for formal Ni valence through modifying the ratio of positively charged RO and negatively charged NiO$_2$ planes, in addition to modifying dimensionality and confinement by varying the layer thickness. 
Ruddlesden-Popper nickelates exhibit a wide range of strongly correlated properties, ranging from spin and charge stripes in antiferromagnetic La$_2$NiO$_4$ \cite{chen1993charge} to the recent stabilization of high-temperature superconductivity in pressurized or epitaxially strained La$_3$N$_2$O$_7$ and La$_4$Ni$_3$O$_{10}$ \cite{sun2023signatures,li2024signature, ko2025signatures}. The $n = 1$ provides an ideal testbed for $s$-NIXS experiments, with single symmetry-equivalent Ni site centered in oxygen octahedral cages echoing those of the perovskite, as well as the stable Ni$^{2+}$ formal valence similar to canonical NiO. 

Fig.~\ref{fig:fig3}a shows the projected structure of La$_2$NiO$_4$, which consists of single two-dimensional layers of NiO$_6$ octahedra stacked along the $c$-axis and separated by LaO rocksalt spacer layers. A nominal Ni$^{2+}$ ion resides in an octahedral coordination within a tetragonal unit cell; due to the presence of the spacer layers,  however, the apical oxygens are displaced away from the central Ni atom along the $c$-axis, thereby lowering the local octahedral symmetry from $O_h$ to $D_{4h}$ and splitting the $e_g$ states (Fig.~\ref{fig:fig3}b). 

Fig.~\ref{fig:fig3}c displays the $s$-NIXS spectra for a wide range of angles $\phi$ within the $(010)$ scattering plane. Direct integration of the spectra between the gray boundaries in Fig.~\ref{fig:fig3}c yields the polar plot shown in Fig.~\ref{fig:fig3}d. The plot reveals an $e_g$-like orbital shape, consistent with the formal high-spin $^3A_2$ ground state of the Ni$^{2+}$ ion. 

Looking in more detail at the spectra displayed in Fig.~\ref{fig:fig3}c, one can observe that the energy of the Ni $M_1$ structure also shifts periodically with angle, in strong contrast to the case of LaNiO$_2$ and LaNiO$_3$ where the Ni $M_1$ peak stays fixed in energy. These energy shifts indicate the presence of crystal fields that split the $s$-NIXS final states of the type $\underline{s}3d^{n+1}$ where $\underline{s}$ indicates the core hole in the $3s$ orbital and $n = 8$ the number of electrons in the ground state of the Ni$^{2+}$ ion \cite{amorese2021selective}.

 Employing a similar approach to analyze the symmetry of the final states as in \cite{amorese2021selective}, we construct a basis using the $S_1$ and $S_2$ spectra. These correspond to maxima of the $e_g$-like orbital shape in Fig.~\ref{fig:fig3}d separated by $90^\circ$, and therefore consist of simple geometric admixtures of $d_{x^2-y^2}$ and $d_{z^2}$ orbital character (see Supplemental Note B for a detailed description). The spectrum at any given angle $\phi$ can be expressed as a linear combination of these two components, such that the corresponding weight coefficients quantify the relative contributions of the $d_{x^2-y^2}$ and $d_{z^2}$ orbital character for each data point in Fig.~\ref{fig:fig3}d. The resulting polar plots of these coefficients are shown in Figs.~\ref{fig:fig3}e~and~\ref{fig:fig3}f. Performing similar analysis for the spectra acquired within the $(001)$ plane (Supplemental Fig. S6) we obtain the polar plots shown in Figs.~\ref{fig:fig3}h~and~\ref{fig:fig3}i.  

Figs.~\ref{fig:fig3}e~and~\ref{fig:fig3}h exhibit a clear $d_{z^2}$ orbital symmetry, while Figs.~\ref{fig:fig3}f~and~\ref{fig:fig3}i are consistent with a $d_{x^2-y^2}$ character. Consequently, using this orbital imaging technique, we can ascribe the spectral component centered at the blue line in Fig.~\ref{fig:fig3}c to transitions into the empty $d_{z^2}$ orbital of the Ni$^{2+}$ ion and the one centered at the red line to transitions into the empty $d_{x^2-y^2}$. The energy difference between those, i.e. 110.7 eV vs. 111.6 eV, thus unveils the tetragonal crystal field splitting, with the $d_{z^2}$ level lower in energy than the $d_{x^2-y^2}$ by 0.9 eV. This value is in good agreement with previous estimates of 0.76 eV based on polarized x-ray (dipole) spectroscopy and cluster calculations of Sr-doped La$_{1.8}$Sr$_{0.2}$NiO$_4$  \cite{schussler2005spectroscopy}. 

\section{Conclusion}

From our results on LaNiO$_3$, we can already appreciate the remarkable capability of $s$-NIXS to probe the orbital landscape of rare-earth nickelates. The pronounced $e_g$ orbital symmetry observed in Figs.~\ref{fig:fig1}c~and~\ref{fig:fig1}d immediately rules out a high-spin $3d^7$ configuration for Ni, indicating that the negative charge transfer energy associated with the high oxidation state of the Ni stabilizes spin-compensated low-energy states. In LaNiO$_2$, our results support a picture of Ni$^{1+}$ $d_{x^2-y^2}$ hole density. 
Although the presence of homogeneously distributed orthogonal domains gives rise to an $e_g$-like orbital in Fig.~\ref{fig:fig2}, the underlying interpretation remains robust when combined with other spectroscopic evidence from XAS and EBSD. We can further exclude the possibility of minor contributions from possible holes in $d_{z^2}$ states because the spectral features associated with non-degenerate $d_{x^2-y^2}$ and $d_{z^2}$ symmetries would be energy-resolved (as in our La$_2$NiO$_4$ measurements), whereas the LaNiO$_2$ spectra exhibit only one spectral peak consistent with a single orbital symmetry. In La$_2$NiO$_4$, $s$-NIXS enables the direct identification of the effective $D_{4h}$ crystal-field splitting between the $d_{z^2}$ and $d_{x^2-y^2}$ components of the otherwise degenerate $e_g$ manifold. In contrast to conventional XAS, where one must rely on cluster-model calculations that explicitly include Ni–O hybridization and bond symmetry to reproduce polarization-dependent spectra \cite{schussler2005spectroscopy}, $s$-NIXS provides a direct and model-independent determination. Not only can we extract the energy splitting, but we also visualize the orbital symmetries of the corresponding states, leaving no ambiguity in their assignment and requiring no further theoretical fitting. This paves the way for applying $s$-NIXS to more complex layered nickelates, such as La$_3$Ni$_2$O$_7$, where orbital imaging could reveal local crystal-field effects and Ni–O hybridization at inequivalent sites of the perovskite blocks \cite{xia2025sensitive}. 

\section{Methods}
\subsection{Sample synthesis}

LaNiO$_3$: Bulk single crystal samples of LaNiO$_3$ were synthesized using high oxygen pressures of 130–150 bar with the optical ﬂoating zone (OFZ) technique as described in Ref. \cite{guo2018antiferromagnetic}. 
\\

LaNiO$_2$: Single crystals of perovskite LaNiO$_3$ were grown by high-pressure OFZ as described in \cite{puphal2023phase,puphal2023synthesis}. 
The as-grown centimeter-sized crystals were oriented with X-ray Laue diffraction and cut into smaller cube-shaped pieces with typical dimensions of approximately 1~mm$^3$ using a wire saw, such that each surface of a cube corresponds to a pseudocubic (100) or (110) plane. 
LaNiO$_2$ crystals  were subsequently obtained via topotactic reduction of the LaNiO$_3$ crystals using CaH$_2$ as the reducing agent. 
Further details of the topotactic reduction procedure and the EBSD analysis of the micro-domains are described in \cite{hayashida2024investigation}.
\\

La$_2$NiO$_4$:
Crystals of La$_2$NiO$_{4+\delta}$ were grown by the floating zone technique in a high pressure mirror furnace (HKZ, Sidre).
Polycrystalline rods were sintered several times at 1100°C in air with intermediate re-grindings.
Floating zone growth was performed with $pO_2$ $\sim$ 25 bar and slow growth rates ($\sim$ 2 mm/h).


\section{Acknowledgments}
This work was supported by the Max Planck Society.
We acknowledge DESY (Hamburg, Germany), a member of the Helmholtz Association HGF, for the provision of experimental facilities.
Parts of this research were carried out at PETRA III.
Data was collected using the XRS spectrometer at the High Resolution Dynamics Beamline P01 provided by DESY Photon Science.
B.H.G. was additionally supported by Schmidt Science Fellows in partnership with Rhodes Trust. This work was supported institutionally and no external funding was received.

\bibliography{sNIXS_nickelates}

\clearpage
\newpage

\renewcommand\thefigure{S\arabic{figure}} 
\renewcommand\thetable{S\arabic{table}} 
\let\oldAA\AA
\renewcommand{\AA}{\text{\normalfont\oldAA}}

\onecolumngrid

\section*{Supplemental Note A: Complete Spectra}

\begin{figure}[h!]
\includegraphics{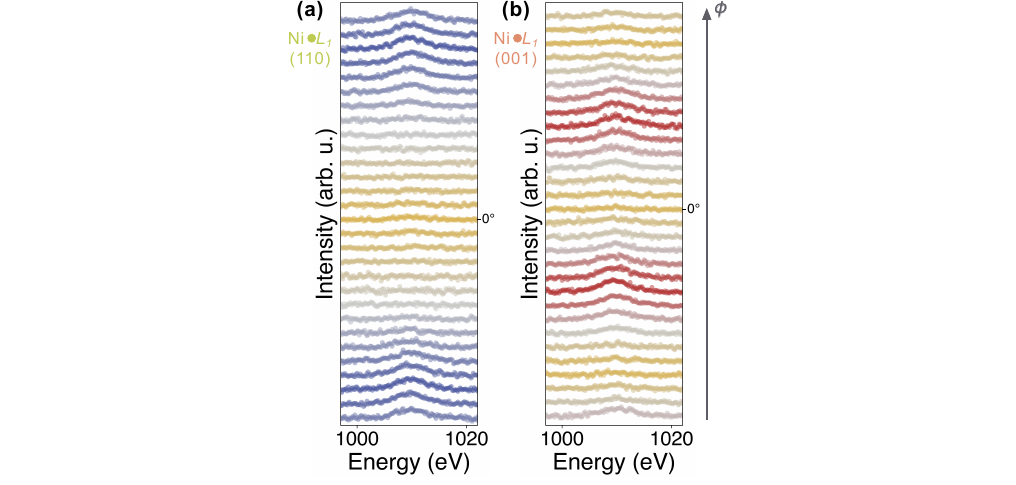}
\caption{\label{SUP_S1} \textbf{Processed spectra of LaNiO$_3$.} (a) Ni $L_1$ edge spectra measured within the $(110)$ plane. (b) Ni $L_1$ edge spectra measured within the $(001)$ plane. The spectra are vertically offset by $\phi$.}
\end{figure}

\begin{figure}[h!]
\includegraphics{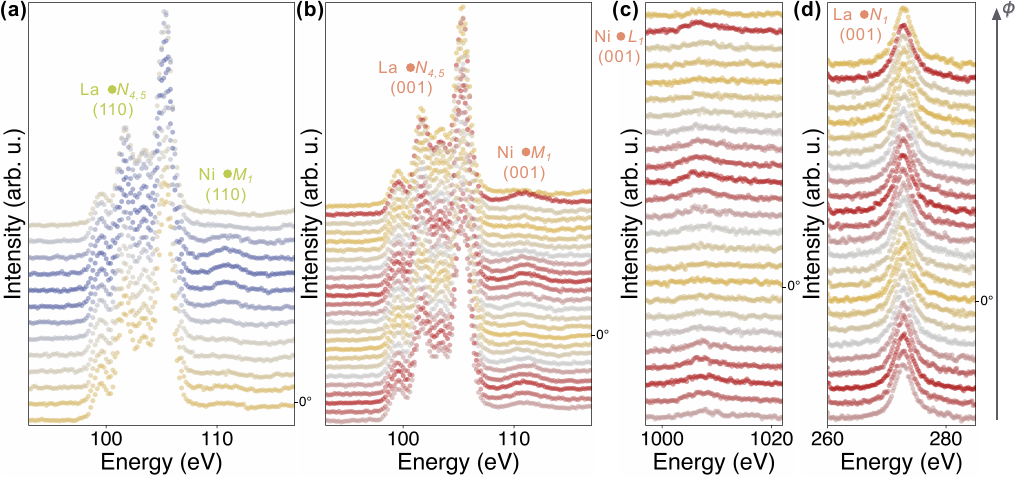}
\caption{\label{SUP_S2} \textbf{Processed spectra of LaNiO$_2$.} (a) Ni $M_1$ edge and La $N_{4,5}$ edge spectra measured within the $(110)$ plane. (b) Ni $M_1$ edge and La $N_{4,5}$ edge spectra measured within the $(001)$ plane. (c) Ni $L_1$ edge spectra measured within the $(001)$ plane. (d) La $N_1$ edge spectra measured within the $(001)$ plane. The spectra are vertically offset by $\phi$.}
\end{figure}

\begin{figure}[h!]
\includegraphics{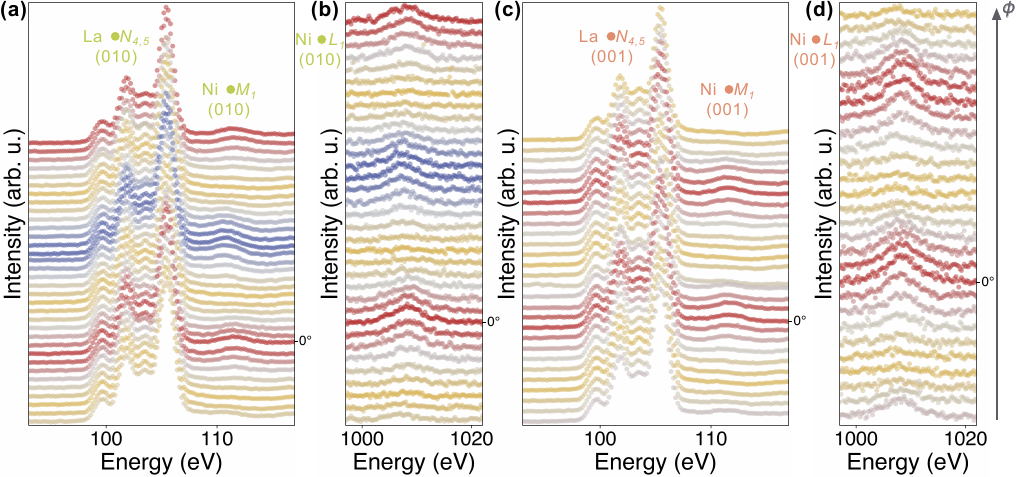}
\caption{\label{SUP_S3}  \textbf{Processed spectra of La$_2$NiO$_4$.} (a) Ni $M_1$ edge and La $N_{4,5}$ edge spectra measured within the $(010)$ plane. (b) Ni $L_1$ edge spectra measured within the $(010)$ plane. (c) Ni $M_1$ edge and La $N_{4,5}$ edge spectra measured within the $(001)$ plane. (d) Ni $L_1$ edge spectra measured within the $(001)$ plane. The spectra are vertically offset by $\phi$.}
\end{figure}

\clearpage
\newpage

\section*{Supplemental Note B: Spectral fitting}

\begin{figure}[h!]
\includegraphics{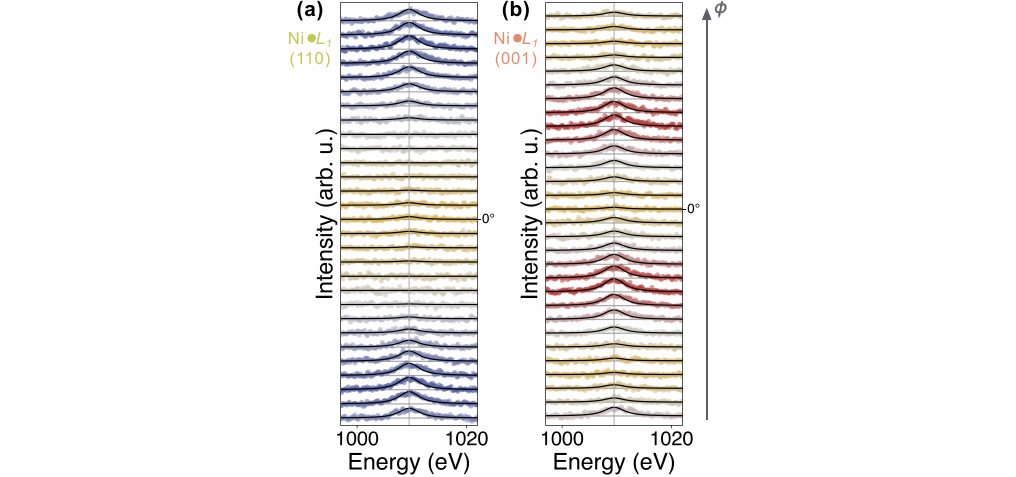}
\caption{\label{SUP_S4} \textbf{Spectral fitting of LaNiO$_3$.} The Ni $L_1$ edge spectra corresponding to the $2s\rightarrow{3d}$ transition were fitted using Voigt profiles. The Gaussian width was fixed to the experimental energy resolution of 1.5 eV, determined from the elastic peaks. The Lorentzian component, representing the core-hole lifetime broadening, was optimized yielding a value of approximately 4 eV. The Voigts are centered at 1009.5 eV. (a) Fitted Ni $L_1$ edge spectra measured within the $(110)$ plane. (b) Fitted Ni $L_1$ edge spectra measured within the $(001)$ plane. The spectra are vertically offset by $\phi$.}
\end{figure}

\begin{figure}[h!]
\includegraphics{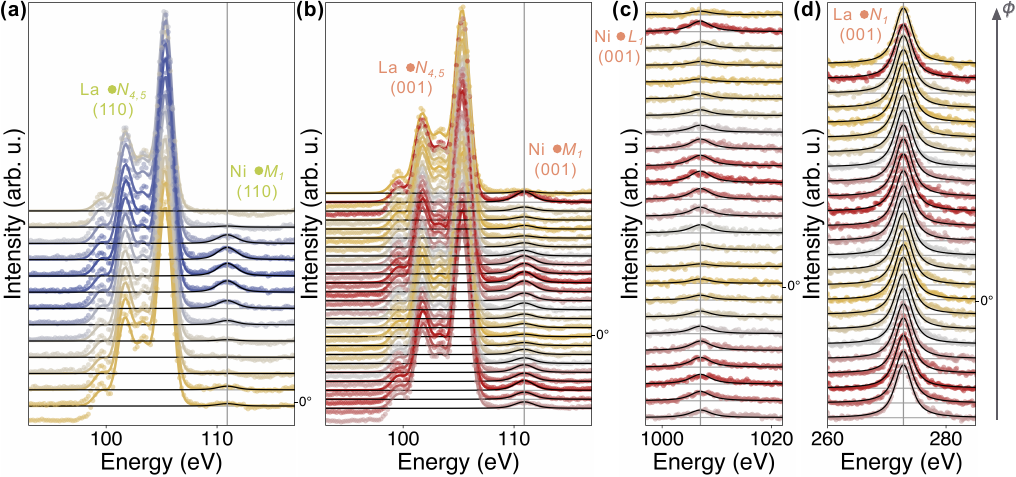}
\caption{\label{SUP_S5} \textbf{Spectral fitting of LaNiO$_2$.} The background of the Ni $M_1$ edge spectra, corresponding to the $3s \rightarrow 3d$ transition, was corrected by applying a linear fit to the pre-edge region of the La $N_{4,5}$ edge and offsetting this fit by a single constant factor across all spectra. The corrected spectra were then fitted using Voigt profiles. To accurately capture the behavior of the $M_1$ peak, the four La $N_{4,5}$ edge features at lower energies were also modeled with Voigt functions. The Gaussian width was fixed to the experimental energy resolution of 1.5 eV, as determined from the elastic peaks. The Lorentzian components were optimized during fitting, yielding a value of approximately 1 eV for the $M_1$ peak, centered at 110.9 eV. The Ni $L_1$ edge spectra were fitted following the procedure described in Supplemental Fig.~S4, with the Voigt profiles centered at 1007 eV. The La $N_1$ edge spectra were fitted using the same approach, with the Voigt profiles centered at 272.8 eV and a Lorentzian broadening of 2.8 eV. (a) Fitted Ni $M_1$ edge and La $N_{4,5}$ edge spectra measured within the $(110)$ plane. (b) Fitted Ni $M_1$ edge and La $N_{4,5}$ edge spectra measured within the $(001)$ plane. (c) Fitted Ni $L_1$ edge spectra measured within the $(001)$ plane. (d) Fitted La $N_1$ edge spectra measured within the $(001)$ plane. The spectra are vertically offset by $\phi$.}
\end{figure}

\begin{figure}[h!]
\includegraphics{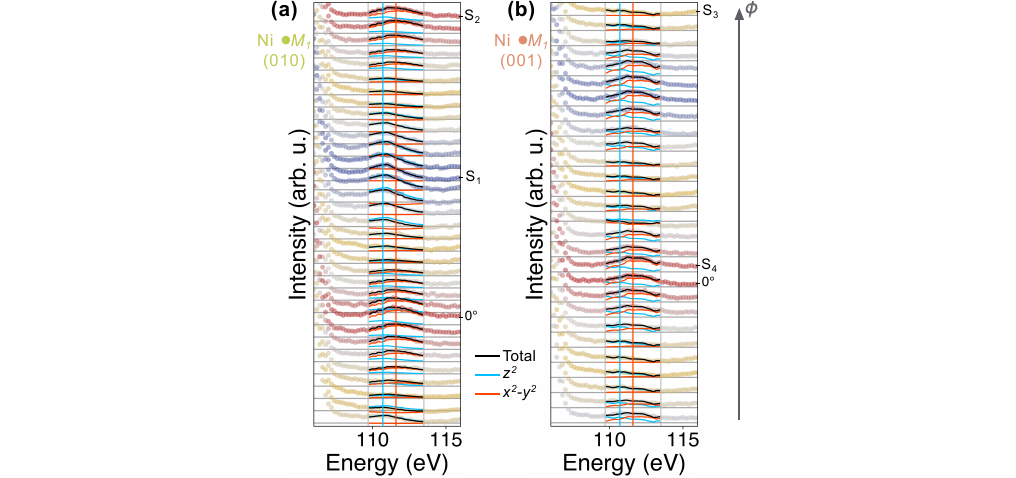}
\caption{\label{SUP_S6} \textbf{Spectral fitting of La$_2$NiO$_4$ as described below.} (a) Fitted Ni $M_1$ edge spectra measured within the $(010)$ plane. Selected $S_1$ and $S_2$ spectra to build the fits basis are indicated. (b) Fitted Ni $M_1$ edge spectra measured within the $(001)$ plane.  Selected $S_3$ and $S_4$ spectra to build the fits basis are indicated. The spectra are vertically offset by $\phi$.}
\end{figure}

\clearpage
\newpage

The Ni $M_1$ edge spectra in La$_2$NiO$_4$ were fitted following a procedure similar to that described in \cite{amorese2021selective}. We constructed a basis using the $S_1$ and $S_2$ spectra, such that the spectrum at any angle $\phi$ can be expressed as a linear combination derived from these states. For the $(010)$ scattering plane, where the spectra are shown explicitly in Fig.~3b of the main text, the $S_1$ spectrum is purely of $d_{z^2}$ character, since the $d_{x^2 - y^2}$ contribution along the $z$ direction is zero. In contrast, the $S_2$ spectrum satisfies:

\begin{equation}
    S_2 = (d_{x^2-y^2})_2 + (d_{z^2})_2 = (d_{x^2-y^2})_2 + \frac{1}{4}(d_{z^2})_1 = (d_{x^2-y^2})_2 + \frac{1}{4}S_1,
\end{equation}

where the subscripts denote the measurement directions and we have used the fact that the ratio between the main lobe and the toroidal (donut-like) lobe in the $d_{z^2}$ charge density is 4. Therefore, we can explicitly quantify the amount of $d_{z^2}$ and $d_{x^2 - y^2}$ orbital character of the spectrum $S$ at $\phi$ by calculating the coefficients $a$ and $b$ from:

\begin{equation}
    S = a(d_{z^2})_1 + b(d_{x^2-y^2})_2,
\label{eq2}
\end{equation}

where $(d_{z^2})_1=S_1$ and $(d_{x^2-y^2})_2=S_2-\frac{1}{4}S_1$. For the $(001)$ scattering plane, where the selected states are indicated in Fig.~3g of the main text, the $S_3$ spectrum is purely of $d_{z^2}$ character, since the $d_{x^2 - y^2}$ contribution along the diagonals within the $(001)$ plane is zero. In contrast, the $S_4$ spectrum satisfies:

\begin{equation}
    S_4 = (d_{x^2-y^2})_2 + (d_{z^2})_2 = (d_{x^2-y^2})_2 + (d_{z^2})_1 = (d_{x^2-y^2})_2 + S_3,
\end{equation}

where we have used the fact that the $d_{z^2}$ contribution is constant within the $(001)$ plane. Therefore, in this case we use Eq.~\ref{eq2} with $(d_{z^2})_1=S_3$ and $(d_{x^2-y^2})_2=S_4-S_3$. In both cases, we solve for $a$ and $b$ employing a standard least-squares method. The resulting decompositions are shown in Fig.~\ref{SUP_S6}. The Ni $L_1$ edge spectra shown in Figs.~\ref{SUP_S3}b~and~\ref{SUP_S3}d were also analyzed; however, the Lorentzian broadening (approximately four times larger than that of the $M_1$ edge) combined with the higher overall noise level, made these data less suitable for studying energy-resolved features.

\begin{figure}[h!]
\includegraphics{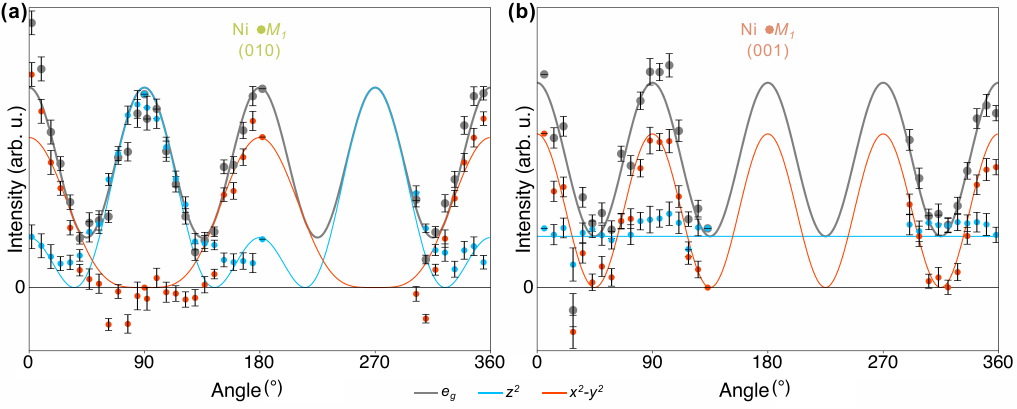}
\caption{\label{SUP_S7} \textbf{Orbital shape analysis in La$_2$NiO$_4$} resulting from energy-resolved M$_1$ edge peak fits around 110.7 eV and 111.6 eV assigned to $z^2$ and $x^2-y^2$ orbitals, respectively, showing unfolded polar plots within the (a) $(010)$ plane and (b) $(001)$ plane. The error bars correspond to the residuals of the fits in Figs.~\ref{SUP_S6}a~and~\ref{SUP_S6}b, respectively.}
\end{figure}

In Fig.~\ref{SUP_S7} we show unfolded representations of the polar plots from Fig.~3 of the main text for the $(010)$ and $(001)$ scattering planes, using the residuals of the fits in Fig.~\ref{SUP_S6} as error bars. Points with negative fitted coefficients were assigned a radius of zero in the polar plots. These artifacts mostly originate from background-correction instabilities related to directional sample dependence. Nevertheless, we chose not to apply any local background adjustments in order to maintain consistency throughout the analysis.

\clearpage
\newpage

\section*{Supplemental Note C: Peak positions}

\begin{table}[h!]
\begin{ruledtabular}
\begin{tabular}{lcc}
\textbf{Sample} & \textbf{Ni $L_1$ edge (eV)} & \textbf{Ni $M_1$ edge (eV)} \\
\hline
LaNiO$_3$ & 1009.5 & not measured \\
LaNiO$_2$ & 1007.0 & 110.9 \\
La$_2$NiO$_4$ &  not resolved & 110.7, 111.6 \\
\end{tabular}
\end{ruledtabular}
\caption{\label{tab1} Measured Ni peak positions for the different compounds.}
\end{table}


\end{document}